\documentclass{aastex63}
\usepackage{graphicx}
\usepackage{CJK}
\usepackage{makecell}
\received{11 Nov 2023}
\revised{15 Jan 2024}
\accepted{06 Feb 2024}
\acceptjournal{ApJ}

\begin{document}
\begin{CJK*}{UTF8}{gbsn}

\title{The Star Formation History in Local Group Galaxies. I. Ten Dwarf Galaxies}

\correspondingauthor{Yi Ren}
\email{yiren@qlnu.edu.cn}

\author[0000-0003-1218-8699]{Yi Ren (任逸)}
\affiliation{Department of Astronomy, College of Physics and Electronic Engineering, Qilu Normal University, Jinan 250200, China}

\author[0000-0003-3168-2617]{Biwei Jiang (姜碧沩)}
\affiliation{Institute for Frontiers in Astronomy and Astrophysics, Beijing Normal University, Beijing 102206, China}
\affiliation{Department of Astronomy, Beijing Normal University, Beijing 100875, China}

\author[0000-0003-1218-8699]{Yuxi Wang (王钰溪)}
\affiliation{Department of Astronomy, College of Physics and Electronic Engineering, Qilu Normal University, Jinan 250200, China} 

\author[0000-0001-8247-4936]{Ming Yang (杨明)}
\affiliation{Key Laboratory of Space Astronomy and Technology, National Astronomical Observatories, Chinese Academy of Sciences, Beijing 100101, China}

\author[0000-0001-7395-1198]{Zhiqiang Yan (闫智强)}
\affiliation{Key Laboratory of Modern Astronomy and Astrophysics (Nanjing University), Ministry of Education, Nanjing 210093, China}



\begin{abstract}
The star formation histories (SFHs) of galaxies provide valuable insights into galaxy evolution and stellar physics. Understanding the SFHs enables the study of chemical enrichment of galaxies, star formation triggered by interactions, and the behavior of various stellar populations. This work investigates the SFHs of ten dwarf galaxies in the Local Group (LG), which spans a wide range of types, masses, luminosities, and metallicities. The analysis is based on our new sample of the member stars in the LG after removing the foreground dwarf stars by the near-infrared color-color diagram and the Gaia astrometric information. The samples include the most complete and pure red supergiants and asymptotic giant branch stars to gain valuable insights into the recent SFHs of the galaxies. The CMD fitting method is introduced to measure the SFH. The Padova isochrones are used to generate initial model CMDs, accounting for photometric errors and completeness through star field simulations to match the completeness and error distributions of the observed CMDs. Subsequently, the SFHs, distance modulus, and metallicity of the ten dwarf galaxies are determined by fitting the CMDs. The results indicate that the star formation rates (SFRs) of dwarf irregulars show a gradual increase, while those of dwarf ellipticals exhibit a gradual decrease from the past to the present. Furthermore, this work shows that the star formation activity in dwarf ellipticals persisted up to 30 Myr ago. A significant increasing feature in the SFH of NGC 6822 reveals star formation activity triggered by an interaction event.
\end{abstract}

\keywords{Massive stars (732); Red supergiant stars (1375); Local Group (929); Dwarf galaxies (416); Stellar evolutionary models(2046); Star formation(1569)}


\section{Introduction} \label{sec:intro}

The Local Group (LG) harbors more than 35 galaxies, encompassing a wide variety of types (including spirals, irregulars, dwarf irregulars, dwarf spheroidals, dwarf ellipticals, etc.), masses, luminosities, and metallicities. It stands as a crucial laboratory for investigating the properties and evolution of different galaxies. In the LG, dwarf galaxies are the dominant population in number. Historically, the definition of dwarf galaxies focused on their size \citep{1971ARA&A...9...35H,1994ESOC...49....3T} and the presence of dark matter halo \citep{1998ARA&A..36..435M}. Although \citet{1994ESOC...49....3T} provided a quantitative definition designating galaxies fainter than $M_{B} = -16$ ($M_{V} = -17$) and more extended (using half-light radius, $r_{1/2}$) than globular clusters as dwarf galaxies, there is no clear boundary between dwarf and other galaxies in the $M_{B}-r_{1/2}$ diagram. Consequently, morphological classification fails to address differences in physical properties between dwarf and other galaxies, and the relationship between dwarf irregular galaxies and dwarf spheroidals/ellipticals remains elusive.

Dwarf galaxies in the LG are considered fragile systems, potentially formed through tidal forces during the Milky Way and Andromeda's early stages or through galaxy collisions. They formed based on the streams of material and dark matter originating from their parent galaxies, evolving under the influence of both internal mechanisms \citep[e.g., star formation, supernova feedback, galactic scale winds, turbulence;][]{1986ApJ...303...39D,1992MNRAS.256P..43E,2005Natur.436..227W} and external processes \citep[e.g., tidal effects, ram pressure, cosmic ultra-violet background, star formation in nearby massive galaxies;][]{2000ApJ...539..517B,2001ApJ...559..754M,2001ApJ...547L.123M,2001AJ....121.2572G,2013ApJ...765...22B}. All these mechanisms collectively shape galaxy formation and give rise to present-day galaxies. So a galaxy's star formation history (SFH) encodes not only the origin and evolution of an individual galaxy but also the whole LG's history.


A galaxy's SFH quantifies the variation in star formation rates (SFRs) with lookback time and metallicity. For star-resolved galaxies in the LG, the SFH can be investigated in the greatest detail and CMD fitting is a suitable and popular method to infer the SFH. Given the stellar isochrones and initial mass function (IMF), the model CMD can be created by considering the photometric error and completeness. The SFR for each isochrone (i.e., SFH) can be solved by comparing the model CMD with the observed CMD \citep{1997NewA....2..397D,1997AJ....114..669A,2009AJ....138..558A,2001ApJS..136...25H,2002MNRAS.332...91D,2012ApJ...751...60D,2013ApJ...775...76D}. Numerous works (e.g., \citealp{1989MNRAS.241..433F,1991AJ....102..951T,1996AJ....112.1438L,1996ApJ...462..684T,1997AJ....114.2514D,1997AJ....114.2527D,1997AJ....113.1001H,1998AJ....115.1869G,1999AJ....118.1657C,1999AJ....118.2229M,2000ApJ...531..804D,2001ApJ...559..225H,2003AJ....126..187D,2005AJ....129.2217B,2007ApJ...668..876Y,2009ApJ...699L..84S,2012ApJ...747..122C,2014ApJ...786...44S}) have investigated SFHs of ten star-resolved dwarf galaxies in the LG (WLM, IC 10, NGC 147, NGC 185, IC 1613, Leo A, Sextans B, Sextans A, NGC 6822, Pegasus Dwarf) through CMD analysis. With the advancement of observational data and computational capabilities, the methodology has evolved from the comparison of observed CMDs to a few isochrones to the synthetic CMDs, accounting for stellar evolution models, photometric completeness, and error distributions. This progress enables the inference of the long time span and high time resolution of SFHs and chemical enrichment histories.

However, differences in photometric data and methods (varied CMD fitting algorithms, different data reduction methods, and different stellar evolution libraries) make it difficult to compare these results. \citet{2014ApJ...789..147W} addresses these issues by analyzing high-quality Hubble Space Telescope (HST) data and obtained SFHs of 40 dwarf galaxies in the LG. This approach enables comparison of the SFHs among different galaxies and average SFHs across various morphological types. For example, \citet{2014ApJ...789..147W} found that the average lifetime SFHs of dwarf spheroidals can be approximated by an exponential decay function, whereas the SFRs of dwarf irregulars gradually increase for lookback ages less than 10 Gyr. Although we omit additional results and discussions from \citet{2014ApJ...789..147W}, these findings underscore SFHs' crucial role in exploring physical property differences among galaxies with varying morphologies. 

Two important considerations arise when inferring SFH using CMD fitting. The first is foreground contamination, as demonstrated by \citet{2007AJ....133.2393M}, which revealed significant contamination by foreground dwarf stars, even for galaxies outside the Galactic plane. Consequently, when comparing the model CMD with the observed CMD, addressing foreground contamination is essential. \citet{2014ApJ...789..147W} chose not to remove foreground contamination from the observed CMD but instead introduced foreground stars into the synthetic CMD using the Milky Way structural model \citep{2010ApJ...714..663D}. This makes the results dependent on the accuracy of the Milky Way structural model. Additionally, the observed CMD should ideally encompass both the oldest and youngest stellar populations to provide a comprehensive understanding of a galaxy's SFH. There are mainly main sequence stars and red giant branch stars (RGBs) in \citet{2010ApJ...714..663D}, the range of the lookback time is approximately from 300 Myr to 14.1 Gyr ($\log~t$ from 8.5 to 10.15). This suggests that the recent SFHs are hardly strictly restricted. 

\citet{2021ApJ...907...18R,2021ApJ...923..232R} proposed the $J-H/H-K$ criteria which is supplemented by the $r-z/z-H$ diagram \citep{2021A&A...647A.167Y} and astrometric information from Gaia DR2/EDR3 to eliminate the foreground dwarf stars. Since the $H$ band is sensitive to surface gravity \citep{2021A&A...647A.167Y}, the $J-H/H-K$ diagram completely based on the near-infrared photometry, is used to remove the foreground dwarf stars on the basis that the intrinsic color indexes of giant and dwarf stars have clear bifurcations in this color-color diagram (CCD) \citep{1988PASP..100.1134B}. \citet{2021ApJ...923..232R} applied the $J-H/H-K$ CCD method to the 12 low-mass galaxies in the LG, resulting in 510,947 pure member stars (42,074 in ten dwarf galaxies, 66,573 in the Small Magellanic Cloud (SMC), 402,300 in the Large Magellanic Cloud (LMC)), including red supergiants (RSGs), asymptotic giant branch stars (AGBs) and RGBs. These member star samples from \citet{2021ApJ...923..232R} encompass various dwarf galaxies and remain almost unaffected by foreground contamination, offering a comprehensive understanding of the SFHs of dwarf galaxies in the LG. Besides, these member star samples include young stellar populations such as RSGs, which can effectively restrict the recent SFHs.

The paper's organization includes Section \ref{sec:CMD Synthesis} discussing CMD synthesis, Section \ref{sec:CMD Fitting} outlining the method for measuring SFH, and Section \ref{sec:Characteristics of SFHs of Different Galaxies} exploring the characteristics of SFHs among different galaxies.

\section{Observed CMDs} \label{sec:Observed CMDs}
The pure member stars catalogs of ten dwarf galaxies form the foundation for constructing the observed CMDs. \citet{2021ApJ...923..232R} collected the $JHK$ images taken with the Wide Field Camera (WFCAM) from 2005 mid September to 2007 August on the 3.8 m United Kingdom Infra-Red Telescope (UKIRT) located in Hawaii \citep{2013ASSP...37..229I}. WFCAM consists of four Rockwell Hawaii-II (HgCdTe 2048 $\times$ 2048) detectors, each covering $13''.65$ on sky. During the observations reported in \citet{2021ApJ...923..232R}, the average seeing varied between approximately $0''.5 - 1''.5$.

Photometric measurements in the $JHK$ bands for these images were conducted using the Cambridge Astronomical Survey Unit (CASU) pipeline, which utilized the source detection software \emph{imcore}\footnote{http://casu.ast.cam.ac.uk/surveys-projects/software-release/imcore}, resulting in the generation of ASCII-format catalogs\footnote{http://casu.ast.cam.ac.uk/surveys-projects/software-release/fitsio\_cat\_list.f/view}.

To refine the dataset, \citet{2021ApJ...923..232R} utilized the stellar classification flags within the catalogs to select point sources and establish the spatial boundaries of the galaxies to avoid contamination from surrounding field stars. Subsequently, the brightness is corrected for extinction \citep{1998ApJ...500..525S,2019ApJ...887...93G}. Finally, $(J-H)_{0}/(H-K)_{0}$ criteria as well as $(r-z)_{0}/(z-H)_{0}$ method and astrometric information from Gaia EDR3 are introduced to remove foreground stars.

Consequently, the resulting pure member star catalogs were utilized in this work to construct the observed $(J-K)_{0}/K_{0}$ CMDs.

\section{CMD Synthesis} \label{sec:CMD Synthesis}
The process of CMD synthesis unfolds as follows. Firstly, isochrones are chosen to form the initial model CMDs. Subsequently, the model CMDs are refined by accounting for photometric errors and completeness across various magnitudes and color indexes.

\subsection{Padova Stellar Isochrones} \label{sec:Padova Stellar Isochrones}
In this work, the web interface code \emph{CMD 3.6}\footnote{http://stev.oapd.inaf.it/cgi-bin/cmd\_3.6} is employed to generate isochrones. The \emph{CMD 3.6} web interface utilizes Padova stellar evolutionary tracks and incorporates various elements into isochrones, including the thermally pulsing AGBs (TP-AGBs) phase, circumstellar dust, initial mass function, and so on for multiple photometric systems. Input parameters for generating isochrones using the \emph{CMD 3.6} web interface are detailed in Table \ref{tab:Input Parameters for Generating ISOs}. Isochrones are set to contain a total mass of $10^6M_{\odot}$ with $\log t\leq8.6$ and $2\times10^6M_{\odot}$ with $\log t>8.6$, while each isochrone point is assigned an absolute number of stars per unit mass based on IMF. Consequently, an initial model CMD is generated from these isochrones. 

\begin{figure}
	\centering
    \includegraphics[scale=0.34]{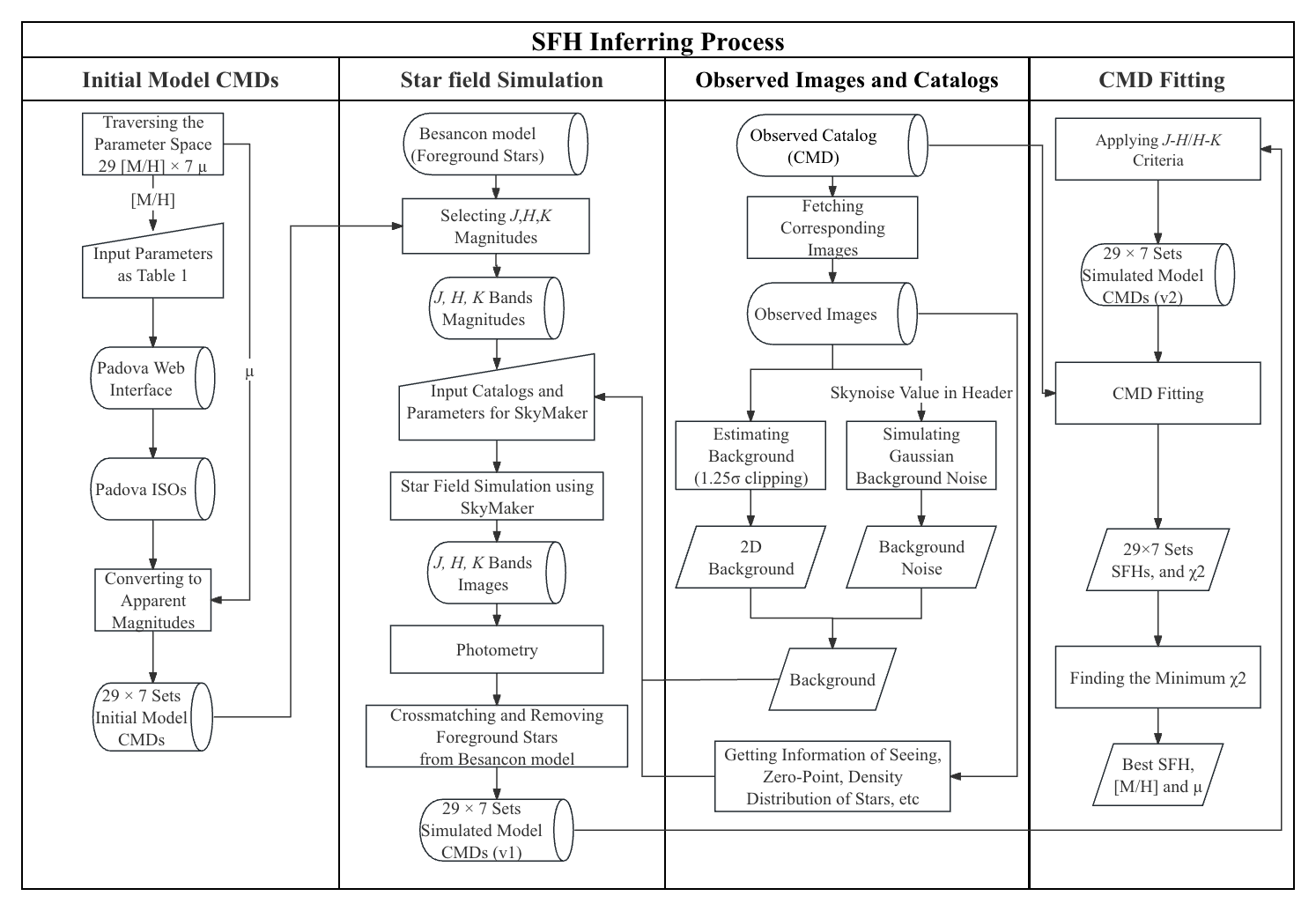}
	\caption{This flowchart shows the SFH inferring process of a galaxy, which consists of three main parts: initial model CMDs generation, star field simulation, and CMD fitting. \label{fig:CMD Fitting Process}}
\end{figure}

\begin{figure}
	\centering
    \includegraphics[scale=0.55]{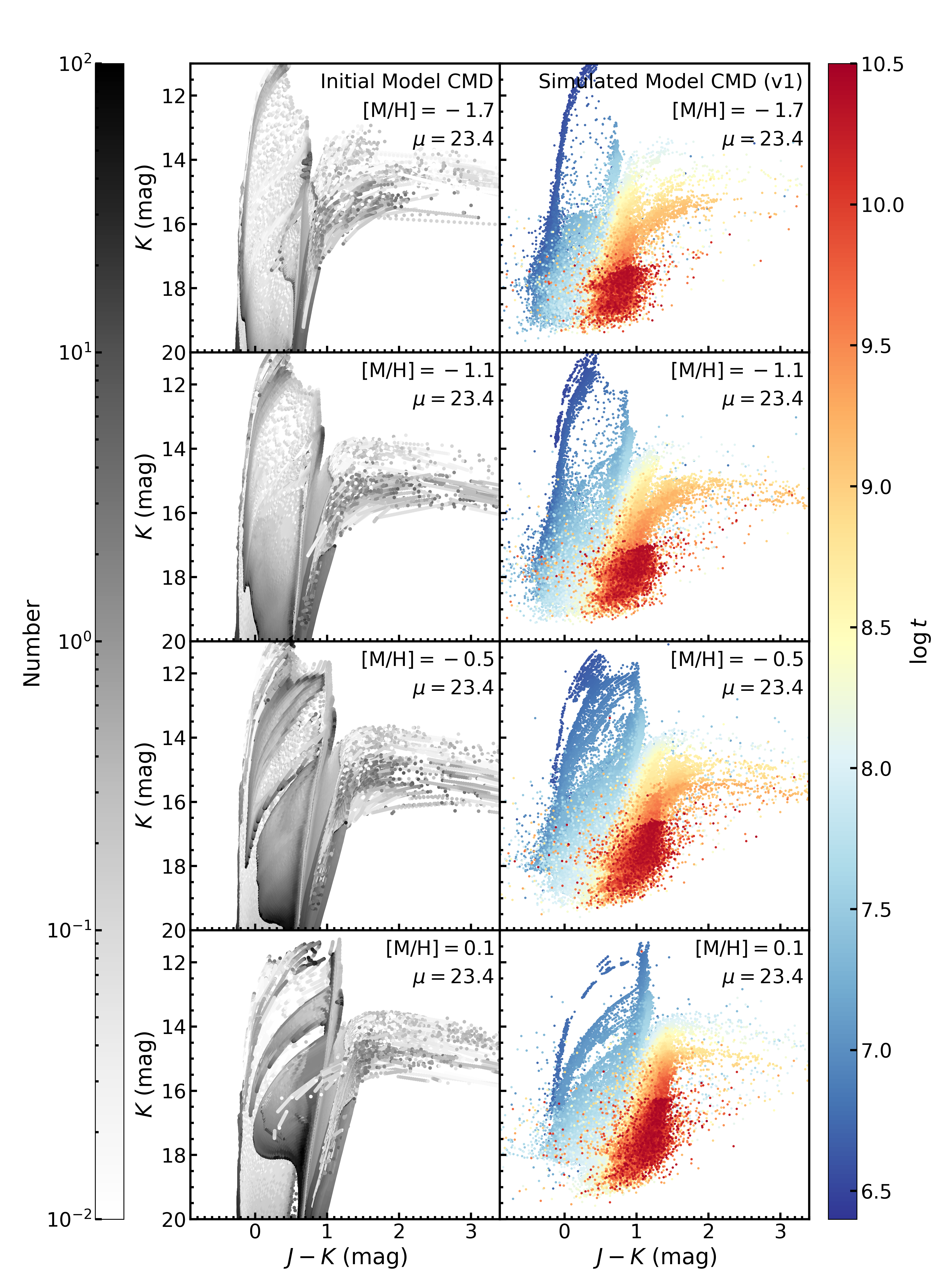}
	\caption{The four panels on the left side display the initial model CMDs at different metallicity. Grayscale of points represents the corresponding number of stars. The four subfigures in the right column show the variation of the simulated model CMDs at a distance modulus of 23.4 under different metallicities. The simulated model CMDs are annotated as `v1' for those without the $J-H/H-K$ criteria applied, and `v2' for those with the criterion applied, to facilitate distinction. The color of the points indicates their age. It can be observed that as the metallicity increases, the simulated model CMD overall shifts towards the red end, and it is also noteworthy that the Tip-RGB (TRGB) becomes brighter in the $K$-band. It should be pointed out that, in order to display more information, the $J-H/H-K$ criteria was not applied to these simulated model CMDs. \label{fig:variation_metallicity}}
\end{figure}

To accommodate the wide range of metallicities and distance moduli found in dwarf galaxies within the Local Group (LG), a grid of initial model CMDs must be constructed. Following the aforementioned approach, the metallicity [M/H] ranges from $-2.2$ to $0.6$ with a resolution of 0.1. The distance modulus ($\mu$) suggested by \citet{2021ApJ...923..232R} serves as the central value ranging between $\pm0.6$ in a step of 0.2. Consequently, a grid consisting of $29\times7$ sets of initial model CMDs is established. During CMD fitting, metallicity and distance modulus are simultaneously determined as free parameters. The process of generating Padova isochrones is visually depicted in the first column of Figure \ref{fig:CMD Fitting Process}. The four panels on the left side of Figure \ref{fig:variation_metallicity} display the initial model CMDs at different metallicities. In the initial model CMDs, the number of stars corresponding to each point is calculated based on the isochrone it belongs to and the total mass set for that isochrone. Since the initial model CMDs are ideal models without photometric errors, points with the same color index and magnitude overlap. Therefore, grayscale is used to represent the number of stars corresponding to each point.

\begin{deluxetable*}{ccc} \label{tab:Input Parameters for Generating ISOs}
	\tablecaption{Input Parameters for Generating isochrones}
	\tablewidth{0pt}
	\tablehead{
		\colhead{Input} & \colhead{Adopted Values} & \colhead{References}
	}
	\startdata
	Evolutionary Tracks & \makecell[c]{PARSEC version 1.2S \\ COLIBRI S\_37 \\ COLIBRI S\_35 \\ COLIBRI PR16} & \makecell[c]{$a$, $b$, $c$ \\ $d$ \\ $e$ \\ $f$, $g$} \\
	Resolution of the Thermal Pulse Cycles & $\mathrm{n}_{\mathrm{inTPC}}=10$ & $h$ \\
	Mass-loss on the RGB & Reimers mass loss with $\eta=0.2$ & $i$, $j$ \\
	Photometric System & $UBVRIJHK$ & $k$, $l$, $m$ \\
	Bolometric Correction & YBC + new Vega & $n$, $o$ \\
	Circumstellar Dust & \makecell[c]{100\% silicate for M stars \\ 100\% amorphous carbon for C stars} & $p$ \\
	Interstellar Extinction & $A_{V} = 0$ & ------ \\
	Long Period Variability & periods from \citet{2021MNRAS.500.1575T} & $q$, $r$ \\ 
	IMF & \makecell[c]{canonical two-part-power law IMF, \\ corrected for unresolved binaries} & $s$, $t$, $u$ \\
	Ages & \makecell[c]{$\log t$ from 6 to 10.5 \\ in a step of 0.02 for $\log t\leq8.6$ \\ in a step of 0.05 for $\log t>8.6$} &\makecell[c]{ \\ ------ \\} \\
	Metallicity & single value & ------ \\
	Simulated Populations with a Total Mass$^*$ & \makecell[c]{$10^6M_{\odot}$ with $\log t \leq8.6$ \\ $2\times10^6M_{\odot}$ with $\log t>8.6$} & ------ \\
	\enddata
	\tablecomments{$^a$\citet{2014MNRAS.445.4287T}, $^b$\citet{2014MNRAS.444.2525C}, $^c$\citet{2015MNRAS.452.1068C}, $^d$\citet{2020MNRAS.498.3283P}\\
	$^e$\citet{2019MNRAS.485.5666P}, $^f$\citet{2013MNRAS.434..488M}, $^g$\citet{2016ApJ...822...73R}, $^h$\citet{2017ApJ...835...77M}\\
	$^i$\citet{1975MSRSL...8..369R}, $^j$\citet{1978AandA....70..227K}, $^k$\citet{2006AJ....131.1184M}, $^l$\citet{1990PASP..102.1181B}\\
	$^m$\citet{1988PASP..100.1134B}, $^n$\citet{2019AandA...632A.105C}, $^o$\citet{2020AJ....160...21B}, $^p$\citet{2006AandA...448..181G}\\
	$^q$\citet{2019MNRAS.482..929T}, $^r$\citet{2021MNRAS.500.1575T}, $^s$\citet{2001MNRAS.322..231K}, $^t$\citet{2002Sci...295...82K}\\
	$^u$\citet{2013pss5.book..115K}\\
	$^*$In fact, this preset value is adjusted according to the estimated number of member stars in each galaxy and selected metallicity. This adjustment aims to decrease the number of simulations and, consequently, the computational core time required, while ensuring that the simulated model CMD comprehensively covers the observed sample distribution.}
\end{deluxetable*}

\subsection{Star field simulation} \label{sec:Star field simulation}
When comparing an initial model CMD with an observed CMD, it is necessary to consider the following factors. The initial model CMDs are the idealized models where all sources are detectable, and there are no photometric errors. In contrast, observed CMDs exhibit dispersion in both color indexes and magnitudes due to photometric errors, along with incompleteness arising from detection limits. 

To simulate the dispersion in an initial model CMD, at a specific point (i.e., color index and magnitude), the covariance matrix can be determined based on the magnitudes, color indexes and their errors given in the observed catalog. The dispersion caused by photometric errors at a specific point in the initial model CMD can then be replicated through random scattering, following a two-dimensional normal distribution with 1$\sigma$ uncertainties in color index, magnitude, and the derived covariance matrix.

For completeness assessment, artificial stars with specific magnitudes and color indexes are added to the original images and subsequently extracted. The completeness of stars is quantified as the ratio of extracted sources to input sources.


This method provides a convenient way to analyze the error distribution and completeness of a CMD, making it suitable for calculating pollution rates \citep{2021ApJ...907...18R,2021ApJ...923..232R,2022Univ....8..465R} and assessing completeness for various stellar populations.

However, observations of dwarf galaxies span different nights over several years. Consequently, accurately simulating the dispersion caused by photometric errors through random scattering becomes challenging. In this work, star field simulation is introduced to consider the CMD dispersion and completeness. The star field simulation faithfully replicates the parameters of each image and simultaneously provides information on error distribution and completeness.

Specifically, for each observation/image, the observation/image information is recorded in the header of the image. Then star field simulation software \emph{SkyMaker} \citep{2009MmSAI..80..422B} is used to simulate the sky, with one of the $JHK$ band magnitudes from the initial model CMD serving as the input catalog, and observation/image information as input parameters. The simulated images are processed using source detection software \emph{imcore}, with arguments matching the recorded values in the header of the corresponding source catalog. Results are cross-matched between the $J$, $H$, and $K$ bands within a $1^{\prime \prime}$ radius, and sources with $\text{N\_Flag} <2$\footnote{The N\_Flag is a flag indicating the number of bands in which the source is identified as a point source \citep{2021ApJ...907...18R}.} are removed. This process yields a simulated model CMD.

Since the star field simulations replicate the same parameters as actual images and use the same source detection program and settings, the simulated model CMDs should exhibit the same dispersion and completeness as the observed CMDs (we have evaluated the effectiveness of star field simulation, as detailed in Section \ref{sec:Evaluation of Star Field Simulation Method}).

An overview of star field simulation is presented in the second column of Figure \ref{fig:CMD Fitting Process}. Further details regarding the key steps of star field simulation will be discussed in Sections \ref{sec:Input Catalog}, \ref{sec:Point Spread Function}, and \ref{sec:Background}.

\subsubsection{Input Catalog} \label{sec:Input Catalog}
The input catalogs contain simulated member stars from the initial model CMDs and simulated foreground stars from Besan\c{c}on model \citep{2003A&A...409..523R}. It should be emphasized here that including the foreground stars in the input catalog simulates the impact of the foreground stars on photometry, providing a more realistic representation of the actual scenario. However, during CMD fitting, the foreground stars in the simulated model CMDs will be removed, ensuring that our results are independent of the galactic structure model.

\begin{figure}
	\centering
    \includegraphics[scale=0.45]{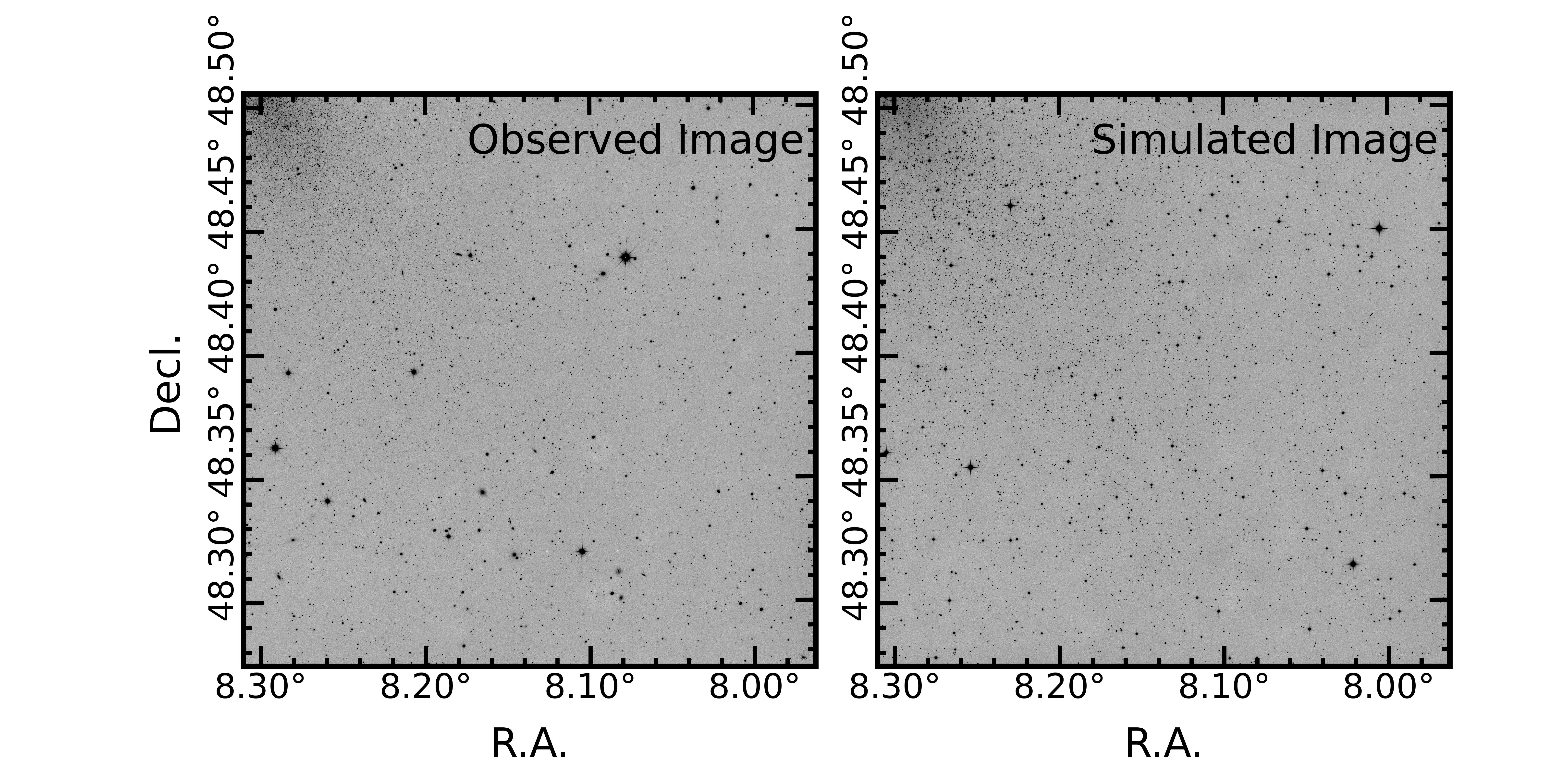}
	\caption{Comparison of an observed NGC 147 image (left panel) and a simulated NGC 147 image (right panel). A portion of NGC 147 is situated in the upper left corner. In the right panel, some bright sources with diffraction spikes are simulated foreground stars from Besan\c{c}on model. \label{fig:Simulated_Image}}
\end{figure}

Additionally, in order to better align with the actual scenario, the position of the simulated foreground stars in the image obeys uniform distribution, with the star field density roughly matching that of recognized foreground stars in \citet{2021ApJ...923..232R}. Similarly, as shwon in Figure \ref{fig:Simulated_Image}, the density distribution of simulated member stars from the initial model CMDs are set according to the distribution of observed member stars from \citet{2021ApJ...923..232R}.

\subsubsection{Point Spread Function} \label{sec:Point Spread Function}
\emph{SkyMaker} uses a tabulated point spread function (PSF) model. The PSF is tabulated at a considerably higher resolution than the final image to accurately capture aliasing effects in conditions of undersampling. The adopted optical transfer function (OTF) of atmospheric blurring is written as:

\begin{equation} \label{eq:PSF}
	\mathrm{OTF}(\|f\|) \propto \exp [-3.442\left(\frac{\lambda\|f\|}{r_0}\right)^{5 / 3} \times\left(1-\left(\frac{\lambda\|f\|}{d_{M_1}}\right)^{1 / 3}\right)],
\end{equation}

\noindent where $\|f\|$ represents the angular frequency, $d_{M_1}$ is the diameter of the primary mirror and taken as 3.8 for 3.8 m United Kingdom Infra-Red Telescope (UKIRT), $\lambda$ is the observation wavelength, and $r_0$ is the \citet{1965JOSA...55.1427F} parameter, which can be calculated as: 

\begin{equation} \label{eq:Fried Parameter}
	r_0 \approx 0.976 \frac{\lambda}{\text {FWHM}}, 
\end{equation}

\noindent where FWHM is the full-width at half maximum of the seeing, with its value same as retrieved from the image header (keyword: SEEING).

\subsubsection{Background} \label{sec:Background}
In background estimation, we adopt a 2-dimensional (2D) background level model and a background noise model rather than remain constant across an image. The Background histogram is clipped iteratively until convergence at $\pm1.25\sigma$ (1.25 is a typical value of the isophotal analysis threshold used in \emph{imcore}) around its median. Subsequently, using a box size of $50\times50$, the background level in each mesh is estimated as the sigma-clipped mode, expressed as $\text{mode} = 2.5 \times \text{median} - 1.5 \times \text{mean}$, following \citet{1996A&AS..117..393B}. For background noise, the root mean square (RMS) is assigned the same value as given in the image header (keyword: SKYNOISE). The final background is determined by adding the background level and background RMS.

\subsubsection{Evaluation of Star Field Simulation Method} \label{sec:Evaluation of Star Field Simulation Method}
\begin{figure}
	\centering
    \includegraphics[scale=0.33]{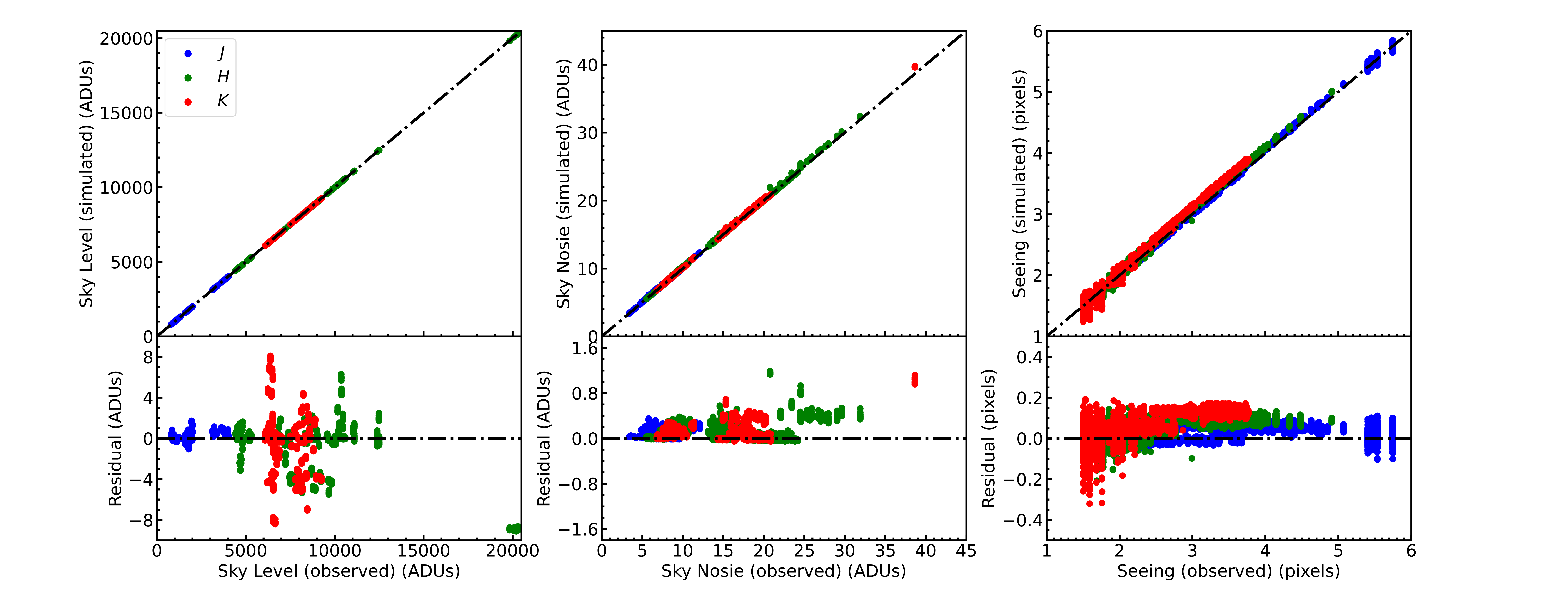}
	\caption{Comparison of parameters for background level (left panel), background noise (middle panel), and average seeing (right panel) between the simulated images and the observed images. The colors indicate $JHK$ respectively. The second row displays residuals, representing the differences between simulated and observed parameters. \label{fig:Star_Field_Simulation_Evaluation}}
\end{figure}

Figure \ref{fig:Simulated_Image} visually displays the effect of star field simulation. To quantitatively evaluate the reliability of star field simulations, we compared simulated images of all galaxies with their original images. Figure \ref{fig:Star_Field_Simulation_Evaluation} shows the comparison of parameters for background level, background noise, and average seeing between the simulated images and the original images, all of which were measured by the \emph{imcore} program. Notably, the background level, background noise, and average seeing in both the simulated and observed images demonstrate strong consistency, indicating that the simulated images can effectively and accurately replicate the background and PSF of the observed images.

\section{CMD Fitting} \label{sec:CMD Fitting}
Since both the simulated model CMDs and observed CMDs share the same completeness and photometric error, these factors are no longer considered during the CMD fitting process to infer SFHs. The observed CMDs are sourced from \citet{2021ApJ...923..232R}, as detailed in Section \ref{sec:Observed CMDs}. To ensure consistency between the observed and simulated model CMDs, the same $J-H/H-K$ criteria employed to remove foreground stars in the observed CMDs are also applied to the simulated model CMDs. The borderline used in $J-H/H-K$ criteria for each galaxy can be found in \citet{2021ApJ...923..232R} in details.

\subsection{Minimization Algorithm} \label{sec:Minimization Algorithm}
For each galaxy, the observed $(J-K)_0/K_0$ CMD is individually matched with $29\times7$ sets simulated model CMDs. Both observed and simulated model CMDs are divided into subregions with a grid size of $0.05\times0.05$, where the $(J-K)_{0}$ ranges from 0 mag to 4 mag and the $K_{0}$ from 10 mag to 22 mag. The number of stars in the $i$th subregion of the model CMDs is expressed as:

\begin{equation} \label{eq:Number of Stars in Model CMD}
	N_{M_{i}} = \sum_j r_j \times \text{ISO}_{i, j}, 
\end{equation}

\noindent where $\text{ISO}_{i, j}$ denotes the number of stars in the $i$th subregion contributed by the $j$th isochrone, $r_j$ is the scale factor. The scale factors, $r_j$, are determined by minimizing the $\chi^2$ value defined as:

\begin{equation} \label{eq:Chi}
	\chi^2=\sum_i \frac{\left(N_{D_i}-N_{M_i}\right)^2}{N_{D_i}},
\end{equation}

\noindent where $N_{D_i}$ is the number of stars in the $i$th subregion of the observed CMD. In Eq. \ref{eq:Chi}, if both $N_{D_i}$ and $N_{M_i}$ are zero, the $i$th subregion do not contribute to the $\chi^2$. If only $N_{D_i}$ is zero, $N_{D_i}$ is replaced with 1 in the denominator \citep{2001ApJS..136...25H}. In this work, adjacent isochrones have fixed intervals of $\log~t$ (i.e., $\Delta \log~t$ takes on the value of 0.02 when $\log~t \leq 8.6$ and 0.05 when $\log~t > 8.6$), therefore, the age interval (i.e., $\Delta t$) occupied by an isochrone is related to $\Delta \log~t$ and expressed as:

\begin{equation} \label{eq:SFH00}
	\Delta \log~t = (\log~t)^{'} \cdot \Delta t.
\end{equation}

\noindent Then, $\Delta t = (\Delta \log~t) \cdot t \cdot \ln 10$, this term represents the time interval occupied by the specific isochrone with an age of $t$. So the star formation rate of the $j$th isochron is $r_{j} \cdot M_{\mathrm{ini}}/\Delta t$, where $M_\mathrm{ini}$ corresponds to the initial total mass of the isochrone, as specified in Table \ref{tab:Input Parameters for Generating ISOs}. Finally, SFH is derived using the formula:

\begin{equation} \label{eq:SFH}
	\text{SFR} (t) = r_{j} \cdot M_\mathrm{ini} / (t \cdot \Delta \log~t \cdot \ln 10).
\end{equation}

Algorithmically, for each simulated model CMD, the Powell method is introduced to optimize the Equation \ref{eq:Chi} and obtain the SFH. The Powell algorithm proceeds by constructing a set of conjugate directions along which it performs one-dimensional minimization. These directions are updated iteratively as the algorithm progresses. In cases of high-dimensional parameters, like in modeling the SFHs, the use of conjugate directions can significantly accelerate convergence speed. This algorithm is valued for its derivative-free approach. Consequently, the Powell algorithm which is built in a Python package called LMfit \citep{2018zndo...1699739N} is employed to determine the SFRs.

For each galaxy, this process yields $29 \times 7$ sets of SFH results. Then, by identifying the minimum $\chi^2$ value among $29 \times 7$ sets CMD fitting, the best-simulated model CMD, complete with specific metallicity and distance modulus, is selected from $29 \times 7$ sets simulated model CMDs. The best simulated model CMDs for nine dwarf galaxies are listed in the second and third columns of Table \ref{tab:The Best Simulated Model CMDs and Corresponding SFHs} (IC 10 is not included). 

\citet{2009AJ....138..558A} developed the IAC-pop code to measure the SFH and chemical enrichment history in a complex stellar population system simultaneously. In comparison, this work do not take into account the chemical enrichment history when calculating the SFH. The IAC-pop code utilizes a $\chi^2$ approach similar to Equation \ref{eq:Chi} and employs a genetic algorithm to minimize $\chi^2$. Additionally, the IAC-pop code requires pre-determined input parameters to account for observational effects, such as photometric errors and incompleteness, on the fitting process. Despite variations in optimization algorithms, approaches to considering observational effects, and the inclusion of chemical enrichment history, both methodologies share a similar approach.

\begin{deluxetable*}{cccccccc}[ht] \label{tab:The Best Simulated Model CMDs and Corresponding SFHs}
	\tablecaption{The Best Simulated Model CMDs for Nine Dwarf Galaxies and Corresponding SFHs}
	\tablewidth{0pt}
	\tablehead{
		\colhead{Galaxy} & \colhead{[M/H]} & \colhead{Distance Modulus} & \colhead{$^{a}$...} & \colhead{$^{b}\mathrm{SFR}_{8.16}$} & \colhead{$\mathrm{SFR}_{8.18}$} & \colhead{$^{c}$...} & \colhead{$^{d}\overline{\mathrm{SFR}}$}\\
		\colhead{} & \colhead{(dex)} & \colhead{(mag)} & \colhead{($M_{\odot}/\mathrm{yr}$)} & \colhead{($M_{\odot}/\mathrm{yr}$)} & \colhead{($M_{\odot}/\mathrm{yr}$)} & \colhead{($M_{\odot}/\mathrm{yr}$)} & \colhead{($M_{\odot}/\mathrm{yr}$)}
	}
	\startdata
	WLM & $-1.1$ & 25.0 & ... & $0.0108^{+0.0048}_{-0.0048}$ & $0.0097^{+0.0042}_{-0.0043}$ & ... & 0.0014 \\
	NGC 147 & $-1.0$ & 24.2 & ... & $0.0001^{+0.0000}_{-0.0000}$ & $0.0120^{+0.0050}_{-0.0047}$ & ... & 0.0145 \\
	NGC 185 & $-1.0$ & 24.1 & ... & $0.0001^{+0.0000}_{-0.0000}$ & $0.0001^{+0.0000}_{-0.0000}$ & ... & 0.0090 \\
	IC 1613 & $-1.1$ & 24.3 & ... & $0.0150^{+0.0064}_{-0.0070}$ & $0.0087^{+0.0040}_{-0.0044}$ & ... & 0.0028 \\
	Leo A & $-1.1$ & 24.5 & ... & $0.0036^{+0.0016}_{-0.0017}$ & $0.0045^{+0.0029}_{-0.0028}$ & ... & 0.0001 \\
	Sextans B & $-0.9$ & 25.8 & ... & $0.0000$ & $0.0027^{+0.0014}_{-0.0017}$ & ... & 0.0002 \\
	Sextans A & $-1.0$ & 25.4 & ... & $0.0070^{+0.0053}_{-0.0058}$ & $0.0023^{+0.0016}_{-0.0015}$ & ... & 0.0003 \\
	NGC 6822 & $-0.9$ & 23.4 & ... & $0.0510^{+0.0207}_{-0.0216}$ & $0.0394^{+0.0151}_{-0.0123}$ & ... & 0.0064 \\
	PegasusDwarf & $-0.6$ & 25.6 & ... & $0.0000$ & $0.0087^{+0.0042}_{-0.0050}$ & ... & 0.0005 \\
	\enddata
	\tablecomments{$^{a}$The SFRs with $6\leq\log t<8.16$ are omitted here.\\
	$^{b}$Subscript 8.16 represents SFR at $\log t=8.16$.\\
	$^{c}$The SFRs with $8.18<\log t\leq10.50$ are omitted here.\\
	$^{d}$For comparison, the lookback time range for calculating the average SFRs are $6.6 \leqslant \log~t \leqslant 10.15$, which are the same as \citet{2014ApJ...789..147W}.\\
	(This table is available in its entirety in machine-readable form.)}
\end{deluxetable*}

\subsection{Two Example SFH Measurements} \label{sec:Two Example SFH Measurements}
\begin{figure}
	\centering
    \includegraphics[scale=0.5]{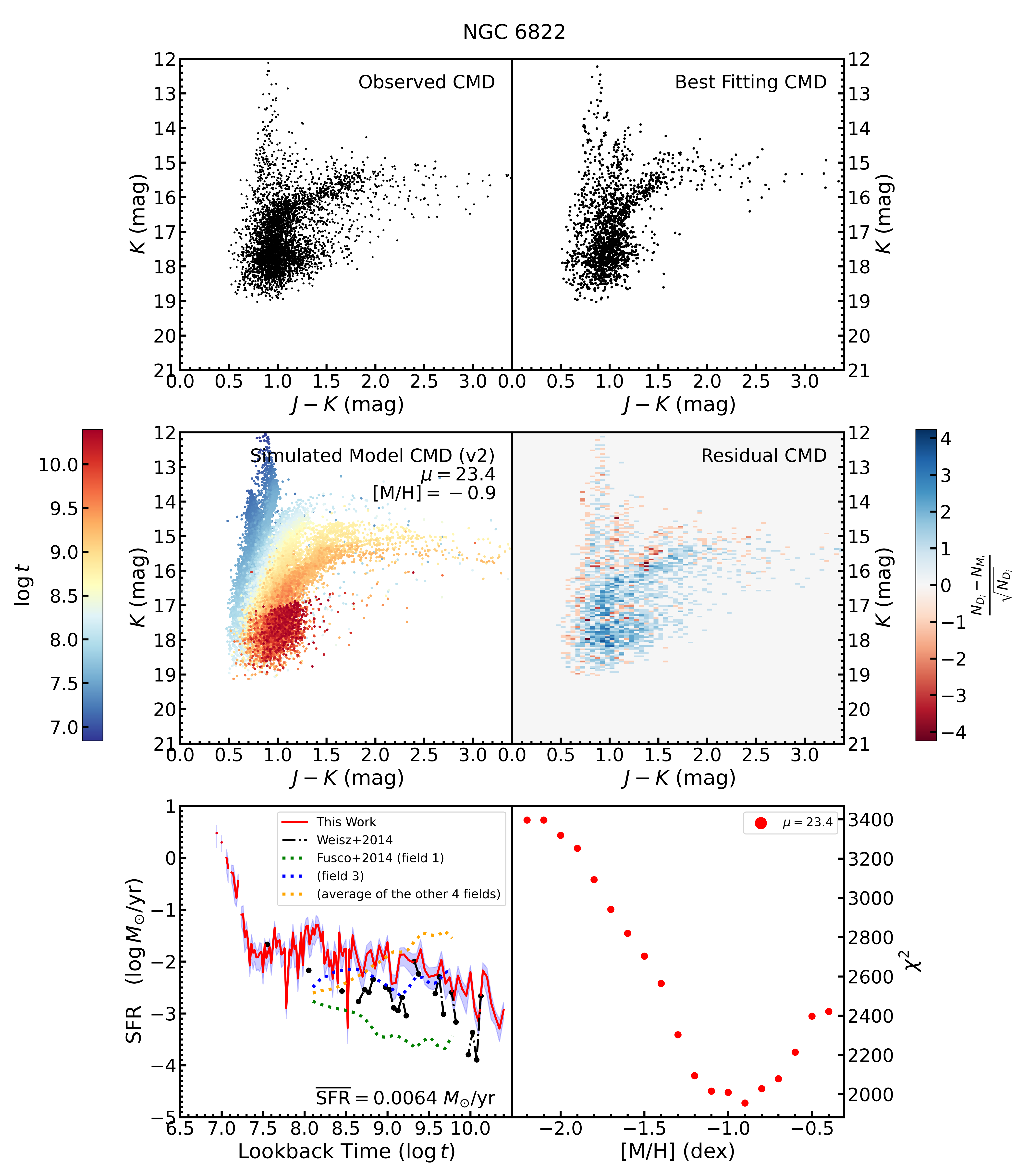}
	\caption{SFH fitting process of NGC 6822. The left upper panel is the observed CMD. The right upper panel is the best fitting CMD. The left middle panel shows the simulated model CMD with determined optimal $\mu$ and [M/H]. The color of dots decodes the stellar age in accord with the color bar. The right middle panel presents the density distribution of residual CMD. The color represents the residual value ($\frac{N_{D_i}-N_{M_i}}{\sqrt{N_{D_i}}}$) in each subregion. In the left lower panel, the best fit SFH is the solid red line. The error envelopes around the best SFH represent the uncertainties based on the 15.865, and 84.135 percentiles of the samples in the posterior marginalized distributions from the emcee. The SFHs of various regions, as reported by \citet{2014AandA...572A..26F}, are depicted using dotted lines. For a consistent comparison within the same coordinate range, the SFH values from \citet{2014AandA...572A..26F} have been scaled by a factor of 50. The SFH obtained by \citet{2014ApJ...789..147W} displays discontinuities in the black dash-dotted line due to instances where the SFRs reach zero at specific $\log~t$ values. In contrast, SFH in this work offers a higher time resolution. The right lower panel is a projection along the $\mu$ axis and only shows the variation of $\chi^{2}$ with [M/H] under the optimal $\mu$. \label{fig:SFH_NGC6822}}
\end{figure}

\begin{figure}
	\centering
    \includegraphics[scale=0.5]{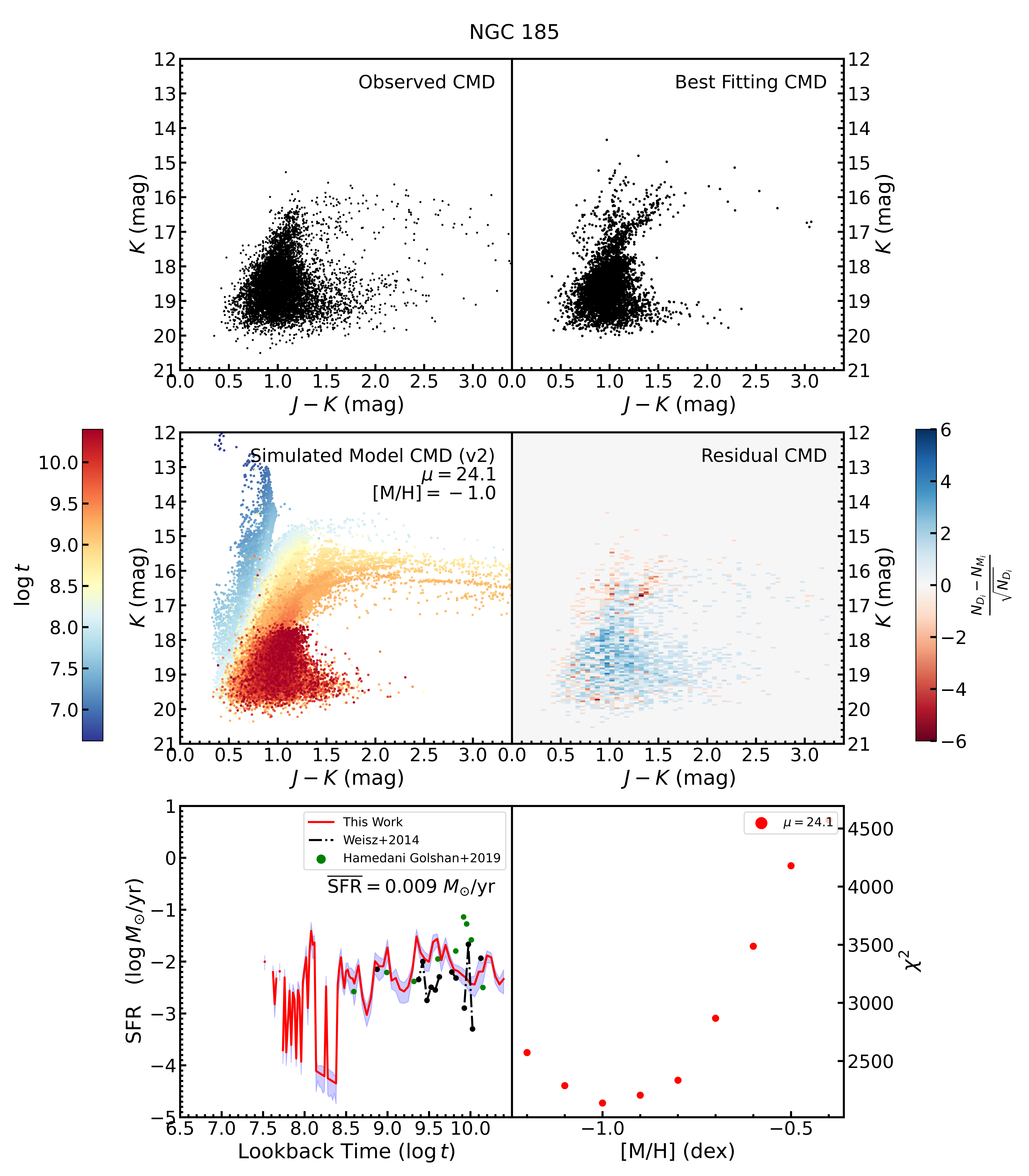}
	\caption{Same as Figure \ref{fig:SFH_NGC6822}, but for NGC 185. \label{fig:SFH_N185}}
\end{figure}

We have chosen NGC 6822 and NGC 185 as representatives of dwarf irregulars and dwarf ellipticals, respectively, to illustrate the CMD fitting process. Figure \ref{fig:SFH_NGC6822} and \ref{fig:SFH_N185} provide a detailed depiction of the SFH measurements process. In these figures:

\begin{itemize}
\item The left upper panels display the observed CMDs.
\item The left middle panels show the simulated model CMDs with optimized $\mu$ and [M/H].
\item The right upper panels reveal the best-fitting CMDs based on the derived SFHs, depicted in the left lower panels.
\item The right middle panels present the 2D density of residuals, showing the difference between the best-fitting CMDs and the observed CMDs.
\item The right lower panels show the method for determining the optimal $\mu$ and [M/H] in all simulated model CMDs. It should be noted that, as this work excludes the chemical enrichment history, the derived best [M/H] should be interpreted as representing the most suitable model grid point. Typically, a thorough analysis encompasses both the SFH and the chemical abundance enrichment history \citep{2009AJ....138..558A,2011ApJ...730...14H}. Indeed, the chemical enrichment history can be obtained based on derived SFH. This information can then be utilized to construct new simulated model CMD for subsequent fitting iterations. Through this iterative refinement and convergence process, an SFH that includes the chemical enrichment history can be obtained. However, this extended process requires integrating stellar evolution models with extremely low metallicity, an aspect not covered in this work.
\end{itemize}

Generally, the structure of best fitting CMDs align with the observed CMDs, and the derived SFHs have similar trends to \cite{2014ApJ...789..147W}. Meanwhile, benefited from the young stellar populations, the SFHs derived in this work span a wide lookback time, ranging from several million years to 10 billion years, offering high resolution at recent lookback times.

However, the 2D density of residual CMDs reveal a systematic difference where the number of RGBs and O-AGBs (Oxygen-rich AGBs) in the best-fit CMD is consistently smaller than in the observed CMD. This feature is also observed in the 2D density of residuals for other nine galaxies. The increasing SFRs of isochrones that match the RGBs and O-AGBs leads to a larger $\chi^{2}$. The issue does not stem from the fitting process but may be attributed to the estimation of the number of stars at TRGB. The observed CMDs show fewer stars near TRGB compared to fainter RGB. On the other hand, in the best-fit CMDs, there are more stars near the TRGB compared to the fainter RGB. So the increasing SFR of isochrones that match the RGBs result in overestimating the number of stars near TRGB, hence a larger $\chi^{2}$. This suggests that the timescale required for stars to evolve to the TRGB while passing through the giant branch is longer than predicted by the model.

\section{Characteristics of SFHs of Different Galaxies} \label{sec:Characteristics of SFHs of Different Galaxies}

To thoroughly compare the results of this work with others, Table \ref{tab:Summary of Different SFH Measurements} summaries the main aspects of each work used for comparison in this paper, including information such as observation and data, main stellar populations in their samples, stellar evolution libraries, and methods for measuring SFHs. Additionally, Table \ref{tab:Overview of Sky Coverage Across Various Works} presents the sky area covered by the samples used in different measurements as in Table \ref{tab:Summary of Different SFH Measurements}.

\begin{deluxetable*}{ccccc}[ht] \label{tab:Summary of Different SFH Measurements}
	\tablecaption{Summary of Different SFH Measurements}
	\tablewidth{0pt}
	\tablehead{
		\colhead{} & \colhead{Observation and Filter} & \colhead{Tracers} & \colhead{Stellar Evolution Library} & \colhead{Methods}
		}
	\startdata
	This Work & \makecell[c]{UKIRT \\ $J$, $H$, $K$} & RSGs, AGBs, RGBs & Padova models & CMD fitting \\
	\citet{2014ApJ...789..147W} & \makecell[c]{HST \\ F555W, F606W, F814W} & $^a$MSs, $^b$RCs, RGBs & Padova models & CMD fitting \\
	\citet{2014AandA...572A..26F} & \makecell[c]{HST \\ F475W, F814W} & MSs, RCs, RGBs & $^c$IAC-star & CMD fitting \\
	\citet{2014ApJ...786...44S} & \makecell[c]{HST \\ F475W, F814W} & MSs, RCs, RGBs & IAC-star & CMD fitting \\
	\citet{2019MNRAS.490.5538A} & \makecell[c]{HST \\ F475W, F814W} & MSs, RCs, RGBs & Padova models & CMD fitting \\
	\citet{2017MNRAS.466.1764H} & \makecell[c]{$^d$NOT, $^e$CFHT \\ $V$, $i$, $J$, $K$} & AGBs & Padova models & $^f$LPVs counts \\
	\enddata
	\tablecomments{$^a$main sequence stars \\
					$^b$red clump stars \\
					$^c$\citet{2004AJ....128.1465A} \\
					$^d$Nordic Optical Telescope \\
					$^e$Canada-France-Hawaii Telescope \\
					$^f$long-period variable stars
	}
\end{deluxetable*}

\begin{deluxetable*}{ccccccc}[ht] \label{tab:Overview of Sky Coverage Across Various Works}
	\tablecaption{Overview of Sky Coverage Across Various Works}
	\tablewidth{0pt}
	\tablehead{
		\colhead{} & \colhead{This Work} & \colhead{$^a$W+2014} & \colhead{$^b$F+2014} & \colhead{$^c$S+2014} & \colhead{$^d$A+2019} & \colhead{$^e$HG+2017} \\
		\colhead{} & \multicolumn{6}{c}{Sky Coverage (arcmin$^{2}$)}
		}
	\startdata
	WLM & 208.63 & 60.82 & ------ & ------ & \makecell[c]{11.36 (inner field) \\ 7.27 (outer field)} & ------ \\
	NGC 147 & 893.29 & 7.26 & ------ & ------ & ------ & 42.25 \\
	NGC 185 & 786.78 & 25.11 & ------ & ------ & ------ & 42.25 \\
	IC 1613 & 892.92 & 14.56 & ------ & 11.36 & ------ & ------ \\
	Leo A & 33.37 & 14.51 & ------ & ------ & ------ & ------ \\
	Sextans B & 56.02 & 7.27 & ------ & ------ & ------ & ------ \\
	Sextans A & 71.82 & 14.56 & ------ & ------ & ------ & ------ \\
	NGC 6822 & 515.58 & 14.56 & 11.36 (each field) & ------ & ------ & ------ \\
	PegasusDwarf & 77.36 & 7.34 & ------ & ------ & ------ & ------ \\
	\enddata
	\tablecomments{$^a$\citet{2014ApJ...789..147W} \\
					$^b$\citet{2014AandA...572A..26F} \\
					$^c$\citet{2014ApJ...786...44S} \\
					$^d$\citet{2019MNRAS.490.5538A} \\
					$^e$\citet{2017MNRAS.466.1764H}
					}
\end{deluxetable*}

The sky area coverage for the sample in this work is based on the surface number density of the initial sample as set by \citet{2021ApJ...923..232R}, defining an area fitted by an ellipse. The semi-major and semi-minor axes of the ellipse are determined as the radius where the stellar number density along each axis drops to the 5$\sigma$ level. As indicated in Table \ref{tab:Summary of Different SFH Measurements}, the sky area coverage of our sample is significantly larger than that in other works (meaning our sample extends to outer regions of the galaxies). This difference in coverage could influence the SFH comparisons. However, it must be pointed out that simply using the ratio of sky area coverage to normalize the results of different measurements into this work is not feasible. This is due to the variability in average SFR across different regions of a galaxy. For instance, \citet{2019MNRAS.490.5538A} analyzed the SFHs of the central and outer fields of WLM using HST ACS and UVIS observations. They defined the central field as being within 1 kpc of WLM's center and the outer field as 1.4 kpc from the center. Their results indicate a strong age gradient in WLM.

\begin{figure}
	\centering
    \includegraphics[scale=0.5]{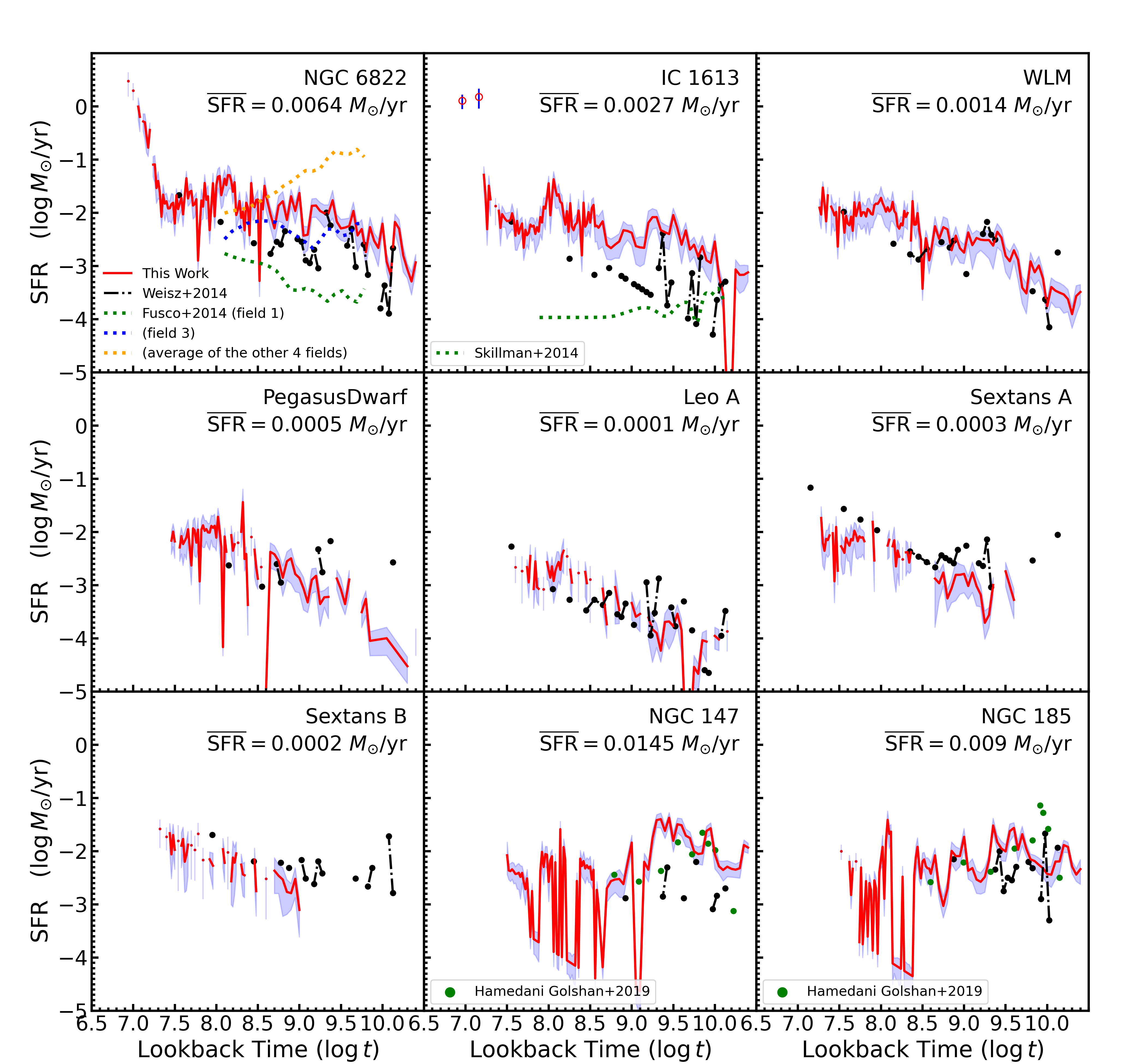}
	\caption{This figure shows the results of SFHs of nine galaxies. The best fit SFHs are the solid red lines. The error envelopes around the best SFHs represent the uncertainties based on the 15.865, and 84.135 percentiles of the samples in the posterior marginalized distributions. In the panel depicting NGC 6822, to ensure a consistent comparison within the same coordinate range, the SFH values sourced from \citet{2014AandA...572A..26F} have been adjusted by a scaling factor of 50. In the subgraph of IC 1613, the measurements of the SFRs represented by two hollow points may originate from the contamination of red giants in the Galactic halo (see Section \ref{sec:IC 1613} for detail). For Sextans A and Sextans B, due to the limited depth of the photometric data \citep{2021ApJ...923..232R}, faint sources (i.e., relatively low mass and relatively old stars) are missing, making it impossible to obtain the early SFHs. This limitation is also reflected in Figure \ref{fig:Cumulative_SFHs_All}.
	\label{fig:SFHs_All}}
\end{figure}

\begin{figure}
	\centering
    \includegraphics[scale=0.5]{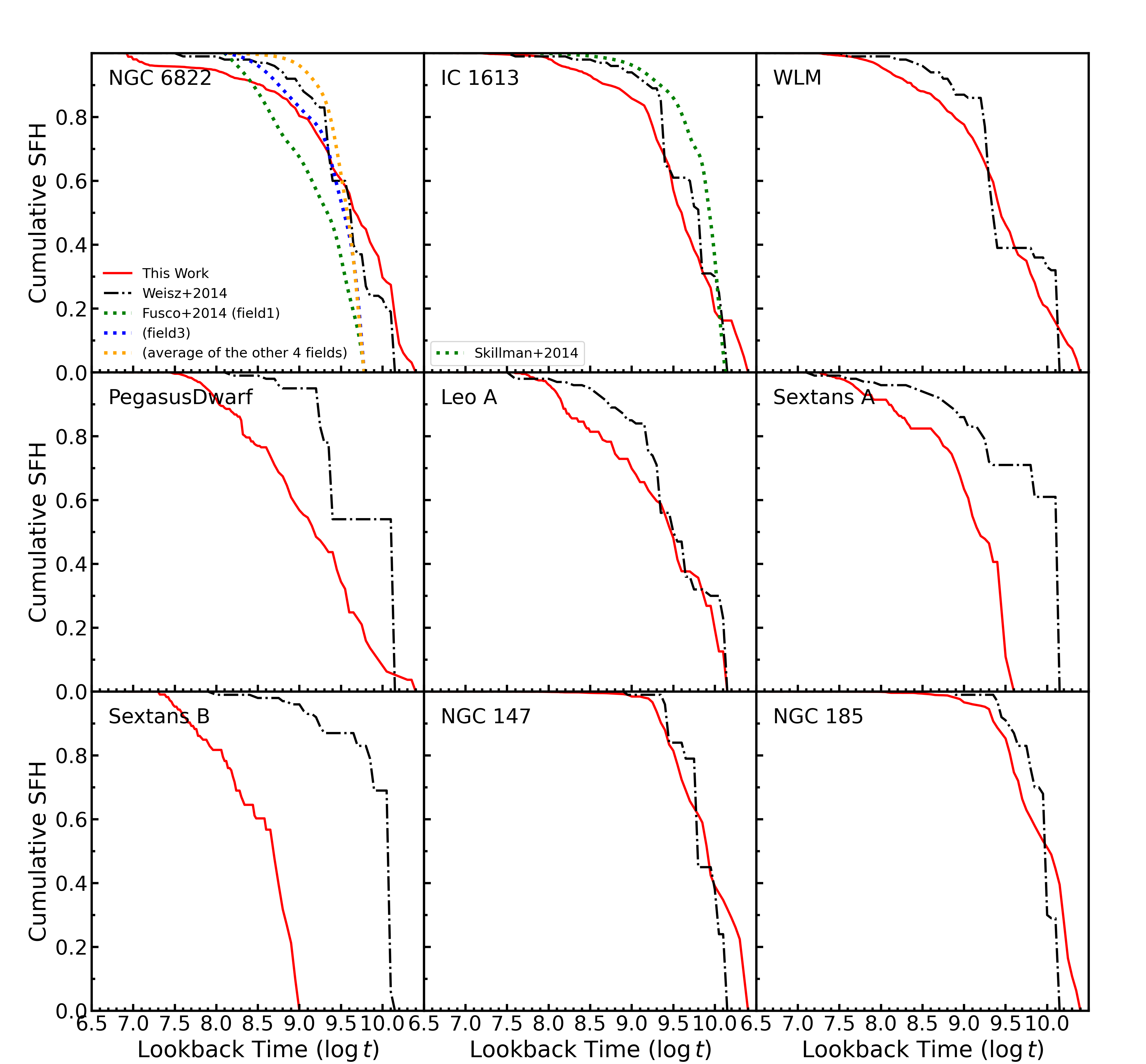}
	\caption{This figure represents the average cumulative SFHs of nine galaxies, respectively. It can be seen that the trends of the cumulative SFHs for individual galaxies in this work are generally consistent with those of \citet{2014ApJ...789..147W}. For NGC 6822, in addition to \citet{2014ApJ...789..147W}, this work reveals the recent SFH of NGC 6822 for $\log~t < 7.5$. During this period, approximately 5\% of the mass of stars in NGC 6822 formed. Since the SFH provided by \citet{2017MNRAS.466.1764H} are discrete points, its cumulative SFH is not calculated. 
	\label{fig:Cumulative_SFHs_All}}
\end{figure}

Figures \ref{fig:SFHs_All} and \ref{fig:Cumulative_SFHs_All} present the SFHs and cumulative SFHs of all galaxies except IC 10, which is discussed separately in Section \ref{sec:IC 10}. Overall, as shown in Figure \ref{fig:SFHs_All} the trends of SFHs derived by \citet{2014ApJ...789..147W} are consistent with this work. However, there are notable differences. Specifically, the SFHs we obtained are characterized by greater continuity, enhanced time resolution, and an extension into more recent time.

Among the remaining nine galaxies, NGC 6822, IC 1613, WLM, Leo A, Sextans A, and Sextans B, PegasusDwarf are dwarf irregulars, NGC 147 and NGC 185 belong to dwarf ellipticals. It is worth noting that PegasusDwarf is a transition galaxy (dIrr/dSph). Because its SFH is similar to dwarf irregulars, we classify it as a dwarf irregular galaxy. The unique characteristics of the SFHs of these two types of galaxies are discussed separately in Sections \ref{sec:Dwarf Irregular Galaxies} and \ref{sec:Dwarf Elliptical Galaxies}.

\subsection{Dwarf Irregular Galaxies} \label{sec:Dwarf Irregular Galaxies}
Dwarf irregular galaxies are characterized by their abundant gas reservoirs and ongoing star formation. Figure \ref{fig:SFHs_All} illustrates that these galaxies share similar SFH patterns, characterized by a gradual increase in the SFR from the past to the present. This indicates that star formation activities in dwarf irregular galaxies are ongoing, and since dwarf irregulars are gas rich, star formation activities themselves can trigger more star formation. However, the recent SFRs of NGC 6822 and IC 1613 show significant increases, which are discussed in Sections \ref{sec:NGC 6822} and \ref{sec:IC 1613}.

\begin{figure}
	\centering
    \includegraphics[scale=0.55]{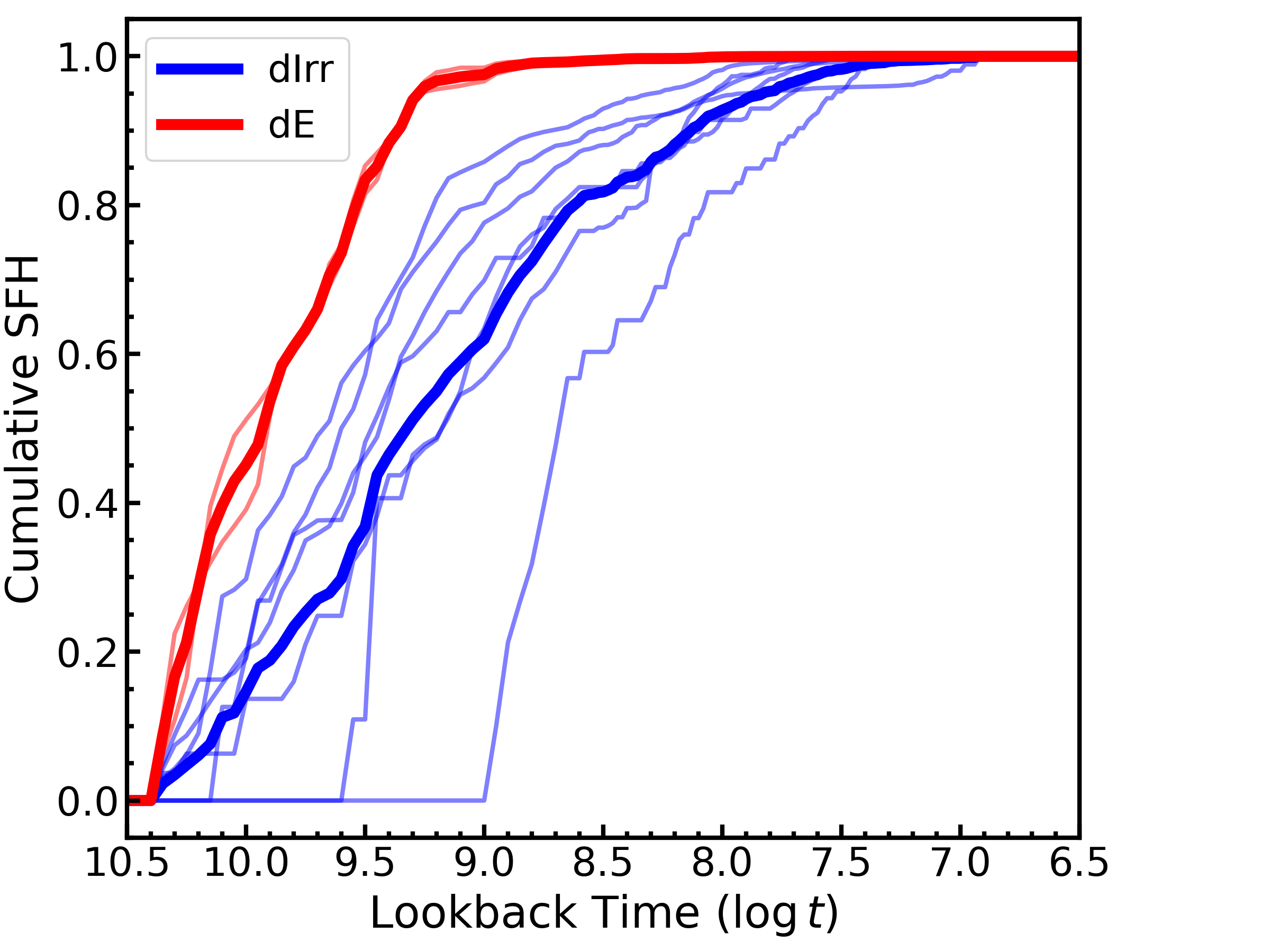}
	\caption{This figure represents the average cumulative SFHs of dwarf irregulars (blue solid line), and dwarf elliptical galaxies (red solid line) with the translucent solid line indicating the individual SFHs in each morphological group. \label{fig:average_SFH}}
\end{figure}

Figure \ref{fig:average_SFH} presents the average cumulative SFHs of dwarf irregulars. It can be seen that dwarf irregulars formed less than $70\%$ of their total stellar mass prior to 1 Gyr, maintain star formation activity toward 10 Myr, and demonstrate an increasing SFR toward the present.

\subsubsection{NGC 6822} \label{sec:NGC 6822}
The recent (from $\sim30$ Myr to the present) SFRs of NGC 6822 show an apparent increase (see Figure \ref{fig:SFH_NGC6822}), and the SFRs are about several to hundreds of times higher than that time prior to $\sim30$ Myr ago.

Such a significant increase in the SFRs may reflect the history of interaction event of NGC 6822, because interactions between galaxies can serve as triggers for star formation.

Previous studies by \citet{1980ApJ...241..125H}, \citet{1996AJ....112.2596G}, and \citet{1999PASP..111..559H} inferred SFH by investigating whole CMD or stellar populations of NGC 6822 and found an enhancement of the SFRs in the last $75\sim200$ Myr (\citealp{1980ApJ...241..125H}, $75\sim100$ Myr ago; \citealp{1996AJ....112.2596G}, $100\sim200$ Myr ago; \citealp{1999PASP..111..559H}, $~100$ Myr ago). \citet{1996AJ....112.2596G} found that the SFRs enhancement in NGC 6822 between 100 and 200 Myr ago was approximately 2 to 6 times.

Moreover, \citet{2000ApJ...537L..95D}, and \citet{2022AJ....164...82P} presented an HI distribution of the NGC 6822. A possible tidal arm with a kinematical timescale of about 100 Myr is found in NGC 6822. This tidal arm is indicative of past interaction events that likely triggered the recent star formation activity observed throughout the galaxy. The $B-R_\mathrm{C}/B$ CMD obtained by \citet{2003ApJ...590L..17K} also supports the interaction event in the history of NGC 6822.

\begin{figure}
	\centering
    \includegraphics[scale=0.65]{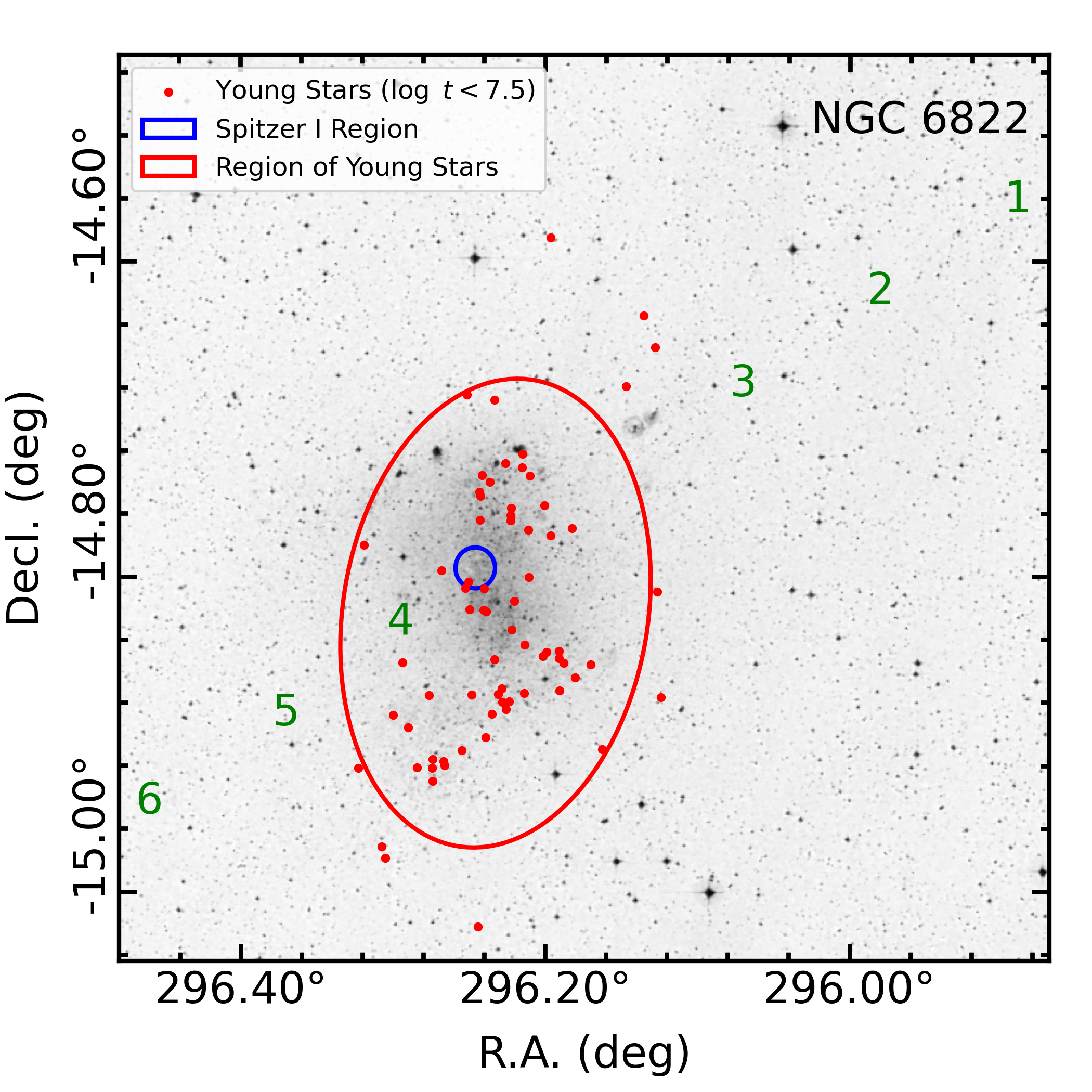}
	\caption{In this figure, the red dots represent young stars with $\log t < 7.5$ with the DSS2 image as the background, and the red ellipse depict the relatively concentrated distribution of these young stars. The red ellipse is manually selected. The semi-major axis of the red ellipse is $0.15^{\circ}$, while the semi-minor axis measures $0.1^{\circ}$. The blue circle delineate the region of Spitzer I with a radius of $0.013^{\circ}$, as specified by \citet{2023arXiv230715704L}.
	\label{fig:region}}
\end{figure}

\citet{2014AandA...572A..26F} selected six fields within the extended HI envelope of NGC 6822 to measure their SFHs based on the HST observations \citep{2012A&A...548A.129F,2012ApJ...747..122C}. The locations of these six fields are indicated by the numbers in Figure \ref{fig:region}, with each region having a size of 11.36 arcmin$^2$. The results show that in the past 500 Myr, the SFRs have increased in fields 1 and 3. In all other fields, the characteristic feature of the last 1 Gyr is a slow decrease in SFR. The SFHs for fields 1 and 3, as well as the averaged SFH of other four fields, are displayed in Figures \ref{fig:SFH_NGC6822} and \ref{fig:SFHs_All}. These works reveal the complex SFH of NGC 6822.

However, it is the first time that such a significant increase in the SFRs of NGC 6822 is measured. To further quantitatively validate this result, we compared it with the findings of \citet{2023arXiv230715704L}. In their work, \citet{2023arXiv230715704L} conducted an imaging survey of the Spitzer I star-forming region in NGC 6822 using the NIRCam and MIRI instruments onboard JWST. They identified 129 young stellar objects (YSOs) and determined a total stellar mass of 1200 $M_{\odot}$ within the Spitzer I region. Assuming a typical YSO formation timescale of $10^{5}$ years, \citet{2023arXiv230715704L} derived the SFR for the Spitzer I region of 0.012 $M_{\odot}$/yr. 

Then we selected member stars with $\log t < 7.5$ in this work, representing the age range associated with a significant increase in SFRs, and illustrated their primary distribution region (young stars region) using red ellipse in Figure \ref{fig:region}. Furthermore, Figure \ref{fig:region} shows the region of Spitzer I depicted by blue circle. We can calculate that the ratio of the areas of the two regions is approximately 88 ($0.15\times0.1/0.013^{2}$). Consequently, we can anticipate that the total recent SFR within the entire region of young stars in NGC 6822 can reach approximately $88\times0.012\approx1.07~M_{\odot}/\mathrm{yr}$. This value is relatively close to the highest measured SFR in NGC 6822, which further strengthens the credibility of our results.

The results presented in this work further corroborate the scenario of an interaction event in the history of NGC 6822 and offer a more precise timeline for the enhancement in its SFRs.

\subsubsection{IC 1613} \label{sec:IC 1613}

In Figure \ref{fig:SFHs_All}, it is evident that IC 1613 exhibits exceptionally high SFRs at $\log~t=6.96$ and $\log~t=7.16$ (also see Table \ref{tab:The Best Simulated Model CMDs and Corresponding SFHs}). These SFRs are nearly a hundred times higher than those at earlier lookback times and contribute approximately $1.6\times10^{6}~M_{\odot}$ to the stellar mass of IC 1613. However, unlike NGC 6822, these two peaks in the SFH of IC 1613 are not continuous and appear as isolated points. To asses the reliability of these unusually high SFRs, we examined the information related to the sources responsible for these peaks, which is summarized in Table \ref{tab:Information on the Two Brightest RSGs in IC 1613}. 
For the UKIRT $JHK$ bands which are listed in \ref{tab:Information on the Two Brightest RSGs in IC 1613} and PS1 $grizy$ bands, not included in Table \ref{tab:Information on the Two Brightest RSGs in IC 1613}, these sources are the brightest. Consequently, we consider the photometry of these two sources to be reliable. However, when estimating distances using the simple relation $r=1/\pi^{\prime \prime}$, we find that the distances to these two sources are approximately 19 kpc and 36 kpc, respectively. While these distances are smaller than the typical scale of the Milky Way \citep[i.e., 50 kpc;][]{2017RAA....17...96L}, it is important to note that Gaia criteria for identifying foreground contamination encompass two conditions: the distance should be less than 50 kpc, and the relative error in the astrometric solution should be smaller than 20\%, which includes $\left| \sigma_{\omega}/\sigma \right| < 20\%$. Therefore, Gaia parallaxes alone may not definitively establish the actual distances of these two sources. Nevertheless, considering the discontinuity of SFH, it is plausible that these two sources are contaminated by red giants in the Galactic halo. Therefore, in the analysis of the SFH of IC 1613, we removed these two points.

\citet{2014ApJ...786...44S} selected an HST ACS field approximately $5'.5$ from the center of IC 1613 (relatively outer region) to investigate its SFH, with their results presented as dotted lines in Figure \ref{fig:SFHs_All}. It is evident that the trend of SFH sourced from \citet{2014ApJ...786...44S} differs from our results, which may reflect the variation in SFH across different regions of IC 1613.

\begin{deluxetable*}{ccccccccc}[ht] \label{tab:Information on the Two Brightest RSGs in IC 1613}
	\tablecaption{Information on the Two Brightest RSGs in IC 1613}
	\tablewidth{0pt}
	\tablehead{
		\colhead{Age} & \colhead{R.A.} & \colhead{Decl.} & \colhead{JMag} & \colhead{HMag} & \colhead{KMag} & \colhead{Parallax} & \colhead{$\mathrm{PM}_{\mathrm{R.A.}}$} & \colhead{$\mathrm{PM}_\mathrm{{Decl.}}$} \\
		\colhead{($\log t$)} & \colhead{(deg)} & \colhead{(deg)} & \colhead{(mag)} & \colhead{(mag)} & \colhead{(mag)} & \colhead{(mas)} & \colhead{(mas/yr)} & \colhead{(mas/yr)}
	}
	\startdata
	7.16 & 16.29002 & 2.20714 & $14.923\pm0.01$ & $14.23\pm0.01$ & $14.063\pm0.01$ & $0.05341\pm0.08527$ & $0.16297\pm0.10697$ & $0.05615\pm0.06921$ \\
	6.96 & 16.25694 & 2.14422 & $13.981\pm0.01$ & $13.294\pm0.01$ & $13.108\pm0.01$ & $0.0275\pm0.05516$ & $0.04075\pm0.06453$ & $0.07342\pm0.04575$ \\
	\enddata
\end{deluxetable*}

\subsubsection{IC 10} \label{sec:IC 10}
IC 10 is known as a starburst dwarf irregular galaxy. In this work, as shown in the extended version of Figure \ref{fig:SFH_NGC6822}, the residual CMD indicates a significant mismatch between the observed CMD and best fit CMD, so IC 10 is not included in further analysis. The Galactic latitudes of IC 10 is $-3^{\circ}$, hence a large foreground extinction $A_{V}\sim4.3$. The large and inhomogeneous extinction can result in such a significant mismatch. This problem would be solved by building a 2D extinction map and performing a more accurate extinction correction for IC 10.

\subsection{Dwarf Elliptical Galaxies} \label{sec:Dwarf Elliptical Galaxies}

Dwarf ellipticals contain little gas. This characteristic is evident in the SFHs of the dwarf elliptical galaxies NGC 147 and NGC 185 of Figure \ref{fig:SFHs_All}. Over time, due to the depletion of gas, the SFHs of NGC 147 and NGC 185 have progressively declined, and they now comprise nearly no young stars such as RSGs, as depicted in Figure \ref{fig:SFH_N185}. \citet{2017MNRAS.466.1764H} investigated the SFHs of NGC 147 and NGC 185 using the method of counting LPVs, with their results displayed in Figure \ref{fig:SFHs_All}. These LPVs are low and intermediate mass AGBs, which also serve as tracers in our sample. Therefore, the SFHs prior to 0.3 Gyr measured by \citet{2017MNRAS.466.1764H} align with those found in this work. When compared to \citet{2014ApJ...789..147W}, as demonstrated in Figure \ref{fig:Cumulative_SFHs_All}, it becomes apparent that NGC 147 and NGC 185 exhibited star formation activity as recent as 30 Myr ago, evidenced by the relatively young stellar populations in this work restricting the SFHs.

\subsection{Average Star Formation Rates Versus HI Mass} \label{sec:Average Star Formation Rates Versus HI Mass}

\begin{figure}
	\centering
    \includegraphics[scale=0.6]{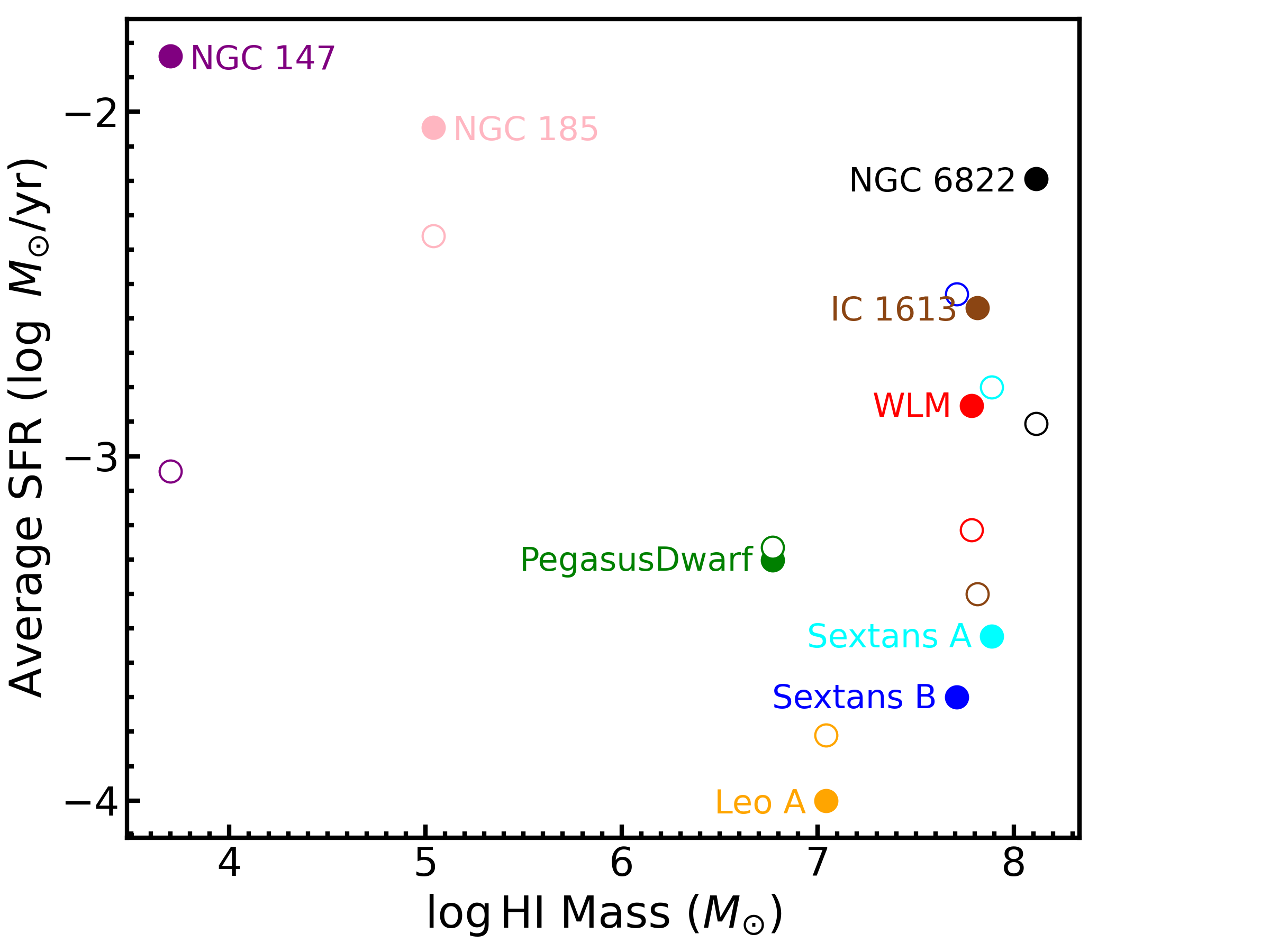}
	\caption{Relation between the average SFRs and the neutral hydrogen masses of galaxies. The average SFRs from this work are denoted by solid circles, and the average SFRs from \citet{2014ApJ...789..147W} are denoted by hollow circles. The masses of neutral hydrogen in galaxies are taken from \citet{2012AJ....144....4M}. \label{fig:SFR_HI}}
\end{figure}

Due to the close relation between star formation activities and galaxies' gas content, we investigated the correlation between average SFRs and neutral hydrogen masses as shown in Figure \ref{fig:SFR_HI}. 

Firstly, it can be seen that the average SFRs of NGC 6822, IC 1613, WLM, NGC 147, and NGC 185 we obtained is greater than the results obtained by \citet{2014ApJ...789..147W}, because recent star formation activities have contributed to the overall SFRs of the galaxies. Even for NGC 147 and NGC 185, the galaxies with weak recent star formation activities, the contribution of its recent star formation to the average SFR of the galaxy cannot be ignored. For Sextans A and Sextans B, due to the incompleteness of our sample, the measured average SFR is lower than that of \citet{2014ApJ...789..147W}.

Secondly, the average SFRs of dwarf irregulars increase with neutral hydrogen mass, while dwarf ellipticals do not exhibit such a correlation. PegasusDwarf, situated as a transitional galaxy, occupies a position between the regions of dwarf irregulars and dwarf ellipticals in Figure \ref{fig:SFR_HI}. This suggests a connection between the morphological classification of galaxies, their star formation activities, and gas content.

\section{Summary} \label{sec:summary and conclusion}
The paper investigates the SFHs of ten dwarf galaxies in the LG. The primary objective is to obtain a comprehensive understanding of the SFHs of dwarf galaxies in the LG by analyzing the CMDs of RSGs, AGBs, and RGBs. In their work, \citet{2021ApJ...923..232R} employed the $J-H/H-K$ method to remove foreground dwarfs and obtained a sample of pure stellar populations from various dwarf galaxies, including RSGs, AGBs, and RGBs. These CMDs of member stars remain unaffected by foreground contamination, providing valuable insights into the SFHs of dwarf galaxies in the LG, particularly focusing on the recent SFHs.

To infer the SFHs, Padova isochrones are selected to generate initial model CMDs. They accounted for photometric errors and completeness by conducting star field simulations, generating simulated model CMDs that matched the completeness and error distributions of the observed CMDs. Subsequently, the CMD fitting method is introduced to obtain the SFHs, distance moduli, and metallicities of the ten dwarf galaxies.

The utilization of high-quality data and advanced methods enables a high-resolution analysis of the SFHs, particularly enhancing our understanding of recent star formation activities in these galaxies.

Dwarf irregulars in our sample set of galaxies exhibit SFHs characterized by a gradual increase in SFRs from the past to the present, whereas dwarf ellipticals have experienced a progressive decline in SFHs. Additionally, the findings from this study indicate that star formation activity in dwarf ellipticals can extend to approximately 30 Myr ago.

IC 10, known as a starburst dwarf irregular galaxy, is excluded from further analysis due to a substantial discrepancy probably caused by significant and inhomogeneous extinction between the simulated model CMD and the observed CMD.

Notably, the recent SFRs of NGC 6822 show an apparent increase, reaching several to hundreds of times higher than the period before approximately 30 Myr ago. This notable SFR surge aligns with an interaction event estimated to have occurred roughly $75-200$ Myr ago.

\acknowledgments{This paper is published in commemoration of the establishment of the Department of Astronomy, Qilu Normal University. We are grateful to Drs. Jing Tang, Bingqiu Chen, Haibo Yuan, and Jian Gao for very helpful discussions and the anonymous referee for very good suggestions. This work is supported by the National Natural Science Foundation of China (NSFC) through grant Nos. 12133002, 12203025, 12303030, and 12373048, Shandong Provincial Natural Science Foundation through project ZR2022QA064, CSST Project CMS-CSST-2021-B03.}

%

\vspace{5mm}
\facilities{}


\software{Astropy \citep{2013A&A...558A..33A},
		  TOPCAT \citep{2005ASPC..347...29T},
		  LMfit \citep{2018zndo...1699739N},
		  dustmaps \citep{2018JOSS....3..695M},
		  SkyMaker \citep{2009MmSAI..80..422B}
          }




\bibliography{paper}{}
\bibliographystyle{aasjournal}



\end{CJK*}
\end{document}